\documentclass[journal, twoside, web]{ieeecolor}

\usepackage{generic}
\usepackage{cite}
\usepackage{amsmath,amssymb,amsfonts}
\usepackage{algorithmic}
\usepackage{graphicx}
\usepackage{textcomp}
\usepackage{url}
\usepackage{xcolor}
\usepackage{subfigure}
\usepackage{makecell}
\usepackage{multirow}
\usepackage{comment}
\usepackage[colorlinks=true,citecolor=blue,urlcolor=blue]{hyperref}
\usepackage{booktabs}
\usepackage{algorithm}
\usepackage{algorithmic}
\usepackage{amssymb}

\def\eg{\textit{e.g.}}
\def\ie{\textit{i.e.}}
\def\etal{\textit{et al.}}

\def\BibTeX{{\rm B\kern-.05em{\sc i\kern-.025em b}\kern-.08em
    T\kern-.1667em\lower.7ex\hbox{E}\kern-.125emX}}
\begin{document}
\title{Contrastive Cross-site Learning with Redesigned Net for COVID-19 CT Classification}
\author{Zhao Wang, Quande Liu, \IEEEmembership{Student Member, IEEE}, and Qi Dou, \IEEEmembership{Member, IEEE}
\thanks{Manuscript received 20 June 2020, revised 15 Aug 2020, accepted 2 Sep 2020. This work is supported by a CUHK start-up research grant and CUHK Shun Hing Institute of Advanced Engineering (project MMT-p5-20). (Corresponding author: Qi Dou.)}
\thanks{Z. Wang is with the College of Information Science and Electronic Engineering, Zhejiang University, Hangzhou, China. (email: kyfafyd@zju.edu.cn)}
\thanks{Q. Liu and Q. Dou are with the Department of
Computer Science and Engineering, The Chinese University of Hong Kong,
Hong Kong, China (emails: \{qdliu; qdou\}@cse.cuhk.edu.hk).}
\thanks{This work was done when Z. Wang did remote internship with CUHK.}
}

\maketitle

\begin{abstract}

The pandemic of coronavirus disease 2019 (COVID-19) has lead to a global public health crisis spreading hundreds of countries.
With the continuous growth of new infections, developing automated tools for COVID-19 identification with CT image is highly desired to assist the clinical diagnosis and reduce the tedious workload of image interpretation. 
To enlarge the datasets for developing machine learning methods,
it is essentially helpful to aggregate the cases from different medical systems for learning robust and generalizable models. 
This paper proposes a novel joint learning framework to perform accurate COVID-19 identification by effectively learning with heterogeneous datasets with distribution discrepancy.
We build a powerful backbone by redesigning the recently proposed COVID-Net in aspects of network architecture and learning strategy to improve the prediction accuracy and learning efficiency. On top of our improved backbone, we further explicitly tackle the cross-site domain shift by conducting separate feature normalization in latent space. Moreover, we propose to use a contrastive training objective to enhance the domain invariance of semantic embeddings for boosting the classification performance on each dataset. 
We develop and evaluate our method with two public large-scale COVID-19 diagnosis datasets made up of CT images.
Extensive experiments show that our approach consistently improves the performanceson both datasets, outperforming the original COVID-Net trained on each dataset by 12.16\% and 14.23\% in AUC respectively, also exceeding existing state-of-the-art multi-site learning methods.

\end{abstract}


\section{Introduction}
\begin{figure}[t]
\centering
\includegraphics[width=0.95\linewidth]{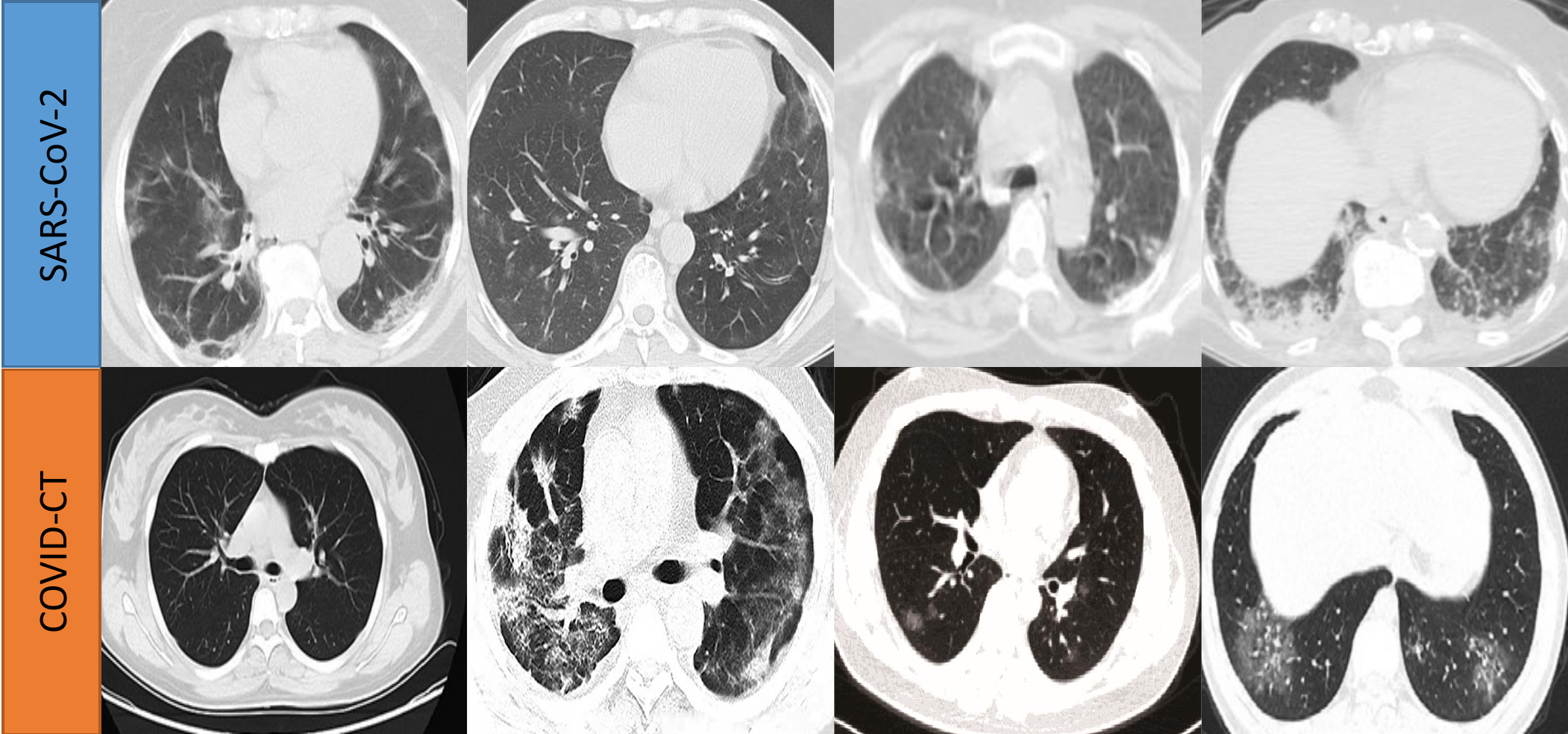}
\vspace{-1mm}
\caption{The CT images of COVID-19 patients from two different clinical centers, showing data heterogeneity on the appearance and contrast.}
\vspace{-4mm}
\label{fig:samples}
\end{figure}

The COVID-19 pandemic, caused by severe acute respiratory syndrome coronavirus 2 (SARS-CoV-2), has lead to a global public health crisis, and continues to spread worldwide. Medical imaging, especially Computed Tomography (CT), has been playing an important role for clinical diagnosis and monitoring of patients with the disease infections~\cite{mei2020ai}.
However, the growth rate of COVID-19 suspicious cases has overloaded the public health service capacity and manifested shortage of trained radiologists. Therefore, developing effective computational methods for automated COVID-19 CT image analysis is highly demanded towards improving the diagnosis outcomes and patient management, as well as helping clinicians on tedious image interpretation workload for releasing their precious time which can otherwise be dedicated to more urgent things on the frontline. 

A considerable amount of data-driven methods have been rapidly developed within this scenario, where the high accuracy is typically attributed to a collected large-scale training database~\cite{javaheri2020covidctnet,xiayong2020covid,fuhuazhu2020infnet}, however, this is difficult to generally achieve in practice.
Instead, to mitigate the insufficiency of single-site data amount, aggregating the CT imaging data from different hospitals is desired for establishing a cross-site learning scheme.
For instance, Di~\etal~\cite{di2020hypergraph} proposed a hypergraph model with multi-site pneumonia data to achieve rapid identification of COVID-19 cases. Wang~\etal~\cite{wang2020covid} developed COVID-Net using data collected from different repositories to build an accurate deep learning classifier for X-Ray images.
However, so far, a major limitation of these works is their negligence of the data heterogeneity across different clinical centers with various imaging conditions (\eg, scanner vendors, imaging protocols, etc). 
As illustrated in Fig.~\ref{fig:samples}, the CT slices of COVID-19 patients from two different public datasets present apparently different image contrasts.
This could potentially affect the model ability to extract robust and general representations as assumed.  
Previous studies on other medical imaging applications~\cite{hu2018inter,john2018variable,liu2020msnet} have frequently observed that straight-forward joint learning with such heterogeneous datasets only brings limited improvement, or even sometimes underperforming individual models trained on a single dataset.

To address this real-world challenge, we propose a novel joint learning framework for accurate identification of COVID-19 CT images by effectively combing different data sources with distribution heterogeneity tackled.
First, we redesign the recent state-of-the-art COVID-Net~\cite{wang2020covid} from aspects of network architecture and learning strategies to boost its computational efficiency and recognition performance. 
Moreover, on top of our new backbone, we conduct effective joint learning to fully exploit the benefit of combining multiple datasets. 
Specifically, our framework employs domain-specific batch normalization layers which enable to conduct the feature normalization and estimate internal feature statistics for each site separately. Importantly, we further propose a contrastive learning objective to explicitly regularize the latent semantic feature space being category sensitive while domain invariant.
We evaluate the effectiveness of our approach using two public COVID-19 CT classification datasets. 
Extensive experiments show that our approach consistently outperforms single-site training models, straight-forward joint learning, as well as existing state-of-the-art multi-site learning methods, on both the datasets.
Our main contributions are summarized as follows:

\begin{itemize}
	\item We redesign the COVID-Net~\cite{wang2020covid} (originally developed for X-Ray) in aspects of network architecture and learning strategy to improve the computation efficiency and prediction accuracy for COVID-19 CT images.
	
	\item We propose a novel joint learning framework to improve the COVID-19 diagnosis by effectively learning from heterogeneous datasets, in which we conduct separate feature normalization to tackle the inter-site data discrepancy and propose a contrastive objective to explicitly promote more robust semantic representations. 
	
	\item Extensive experiments with two public datasets show that our method consistently and significantly improves the classification performance on both datasets.
	Code is available at: \url{https://github.com/med-air/Contrastive-COVIDNet}.
\end{itemize}

The reminder of the paper is arranged as follows. We review the related works in Section~\ref{sec:related}, describe our proposed method in Section~\ref{sec:method}, and elaborate the extensive experiments in Section~\ref{sec:experiment}. We then analyze and discuss our work in Section~\ref{sec:discussion} and finally draw the conclusion in Section~\ref{sec:conclusion}.

\section{Related Works}
\label{sec:related}
\begin{figure*}[t]
\centering

\includegraphics[width=\linewidth]{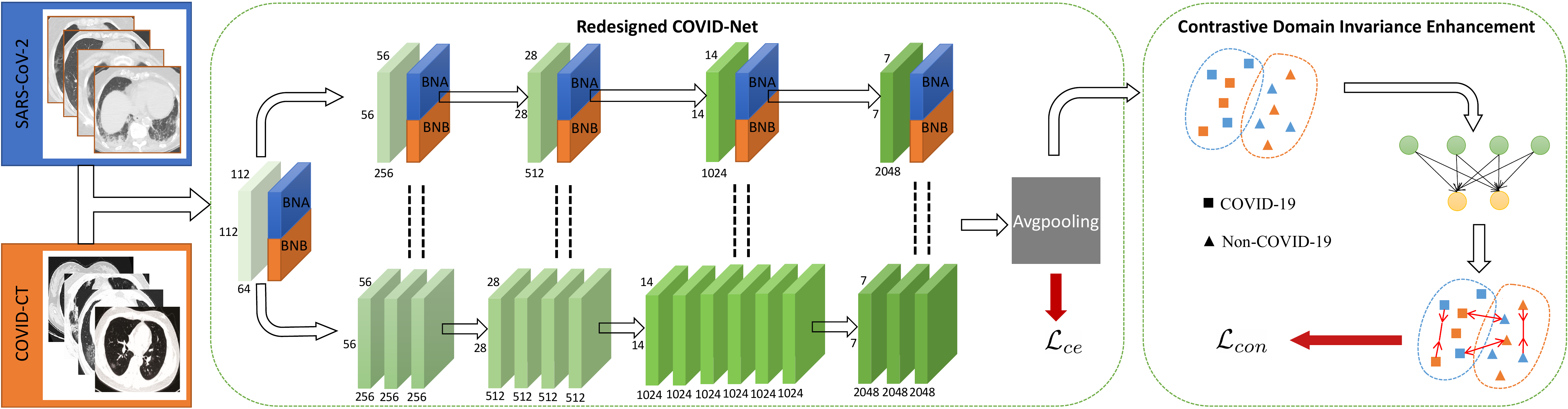}
\vspace{-5mm}

\caption{The overview of our proposed joint learning framework, which redesigns the original COVID-Net as backbone and performs separate feature normalization to tackle the statistical difference of heterogeneous datasets. The proposed contrastive training objective helps to further enhance the domain invariance of semantic embeddings of infected and non-infected cases for boosted diagnosis accuracy on each dataset.}
\label{fig:framework}
\end{figure*}

Many research works have been intensively and rapidly conducted on developing AI methods in responding to COVID-19 global pandemic~\cite{shi2020review}. We hereafter briefly review deep learning approaches for the task of image-level classification for diagnosis which are closely relevant to this paper.

In the beginning, Butt~\etal~\cite{Butt_2020} aimed to establish a screening model for distinguishing COVID-19 pneumonia from those Influenza-A viral pneumonia and healthy cases with chest CT images using ResNet18 with a location-attention mechanism.
Some following-up methods based on transfer learning have been proposed, and most of which used popular existing network architectures, such as VGG \cite{hall2020finding}, ResNet \cite{narin2020automatic, abbas2020classification, farooq2020covid} and DenseNet \cite{li2020covid}. Apostolopoulos~\etal~\cite{Apostolopoulos} relied on MobileNet with its interpretability for helping radiologist to understand how the model prediction was produced.

At the same time, there were new network architectures emerging, carefully designed and validated. Representatively, the COVID-Net \cite{wang2020covid} was tailored for COVID-19 recognition, which achieved a promising accuracy for image-level diagnosis based on chest X-Ray (abbr. CXR). 
Javaheri~\etal~\cite{javaheri2020covidctnet} later designed the CovidCTNet to differentiate positive COVID-19 infections from community-acquired pneumonia and other lung diseases. 
An alternative redesigned framework was based on Capsule Network \cite{afshar2020covidcaps}, aiming to more effectively handle small-scale datasets, which is of valuable significance given the emergency of COVID-19 initial outbreak. The method of Gozes~\etal~\cite{gozes2020rapid} presented a system that can utilize robust 2D and 3D deep learning models, relying on modifying and adapting out-of-the-box AI models and combining them with domain-wise clinical understanding.  
Tang~\etal~\cite{tang2020severity} tackled automated severity assessment (i.e., differentiating non-severe and severe) for COVID-19 based on chest CT images through designed exploration of those identified severity-related features.
Rahimzadeh~\etal~\cite{Rahimzadeh2020} developed a neural network that used concatenation of features from Xception and ResNet50V2 networks, with benefits on recognition performance demonstrated.

With the wide spread of disease, more attentions have been dedicated to joint learning of multiple sites for data sources aggregation. 
For instance, a hypergraph based model~\cite{di2020hypergraph} achieved efficient COVID-19 identification with multi-site pneumonia data; Zhang~\etal~\cite{zhang2020clinically} developed an AI system for COVID-19 diagnosis based on a very large scale dataset (containing about 0.6 million images) which achieved promising performance on several unseen datasets. 
DasAdhikari~\etal~\cite{DasAdhikari2020} combined four datasets based on CT and CXR to study the infection severity of COVID-19. 
Victor~\etal~\cite{victor7effective} used data collected from different repositories~\cite{wang2020covid} for effective COVID-19 screening based on deep learning method.
More broadly speaking, the issues of merging multi-site data have been actively investigated in recent literature on medical image analysis. For instance, Nguyen~\etal~\cite{Nguyen_2019} proposed a novel multi-site learning algorithm to learn different features and aggregate spatial-temporal features through a weighted regularizer based on an integrated multiple heterogeneous dataset.
The deep multi-task learning (MTL) framework~\cite{liao2018deep} could effectively improve the accuracy of  skin lesion classification through the additional context information provided by body location.
Meanwhile, several previous works~\cite{glocker2019machine, van2014transfer} studied the construction of  effective manually generated features and how to design classifiers for medical image analysis tasks across different domains respectively.
The federated learning approach~\cite{li2020multisite} provided private multi-site fMRI analysis through a privacy-preserving pipeline and investigated the federated model’s communication frequency and privacy-preserving mechanisms from various practical aspects.

\section{Methods}
\label{sec:method}
An overview of our framework for COVID-19 diagnosis is illustrated in Fig.~\ref{fig:framework}. In this section, we first describe our model redesign from COVID-Net. We then introduce our joint learning scheme, in which we incorporate separate feature normalization to tackle the cross-site heterogeneity and a contrastive loss to explicitly enhance the domain invariance of latent embeddings for improved classification performance.

\subsection{COVID-Net Redesign for Improved CT Classification}

The starting point of our model is COVID-Net \cite{wang2020covid}, a recent new deep learning architecture for COVID-19 CXR image, that has achieved superior performance over several popular classification networks pretrained on ImageNet. As shown in Fig.~\ref{fig:framework}, the network is composed of two branches, in which the upper branch is a light design with four separate convolutional layers, and the lower branch is composed of blocks with heavier dense connections for representation learning. The skip connection between these two branches are employed for long-range multi-level feature fusion.
However, the COVID-Net~\cite{wang2020covid} was tailored to meet some specific challenges on CXR images in which the lesions are relatively coarse.
Its appropriateness would be changed to a certain extent when applied to CT images where the lesion pattern turns to be more clear, so that presenting richer information to be learned by the model.
In this regard, we aim to build upon the strength of this backbone, while further improving its learning efficiency and classification accuracy from two major complementary angles.

\subsubsection{Network architecture redesign}
One limitation of the original COVID-Net \cite{wang2020covid} is the lack of internal feature normalization layers, which is empirically observed to lead to notable variance of the learned representations across different layers and overall branches. 
As the CT images contain more elaborated patterns, such feature variance will be further amplified if not properly calibrated, which therefore will slow down the the training process and affect the prediction accuracy. 
To address this problem, we incorporate batch normalization~\cite{ioffe2015batch} (BN) layers into the specific components of the network to reduce the internal covariate shift and thus helping improve feature discrimination capability and speed up the convergence rate. 
Importantly, such BN layers are not necessarily beneficial to be naively used as add-on for every single convolution layer.
As the computation blocks in the lower branch contain highly dense short-range connections, adding the BN layers there will significantly increase the parameter scale and decrease training speed. 
As a result, considering the balance between the compution efficiency and stable representation, we add a BN for the initial convolutional layer and a BN after each convolutional layer in the upper branch.

Formally, given $M$-channel feature maps $\textbf{\bf{x}}=\{x_1, \dots, x_M\}$ of a certain layer, the BN obtains the normalized features $\textbf{\bf{y}}=\{y_1, \dots, y_M\}$ by applying affine transformation on the whitened feature maps along each channel $i \in \{1, \dots, M\}$:

\begin{equation}
y_i = \gamma \hat{x_i}+\beta , \quad \text{where} \quad \hat{x_i} = \frac{x_i - \mu_i}{\sqrt{\sigma_i^2 + \epsilon}}, 
\end{equation}
where $\mu_i$ and $\sigma_i^2$  refer to the mean and variance of feature $x_i$; $\epsilon$ is an infinitesimal; $\gamma$ and $\beta$ are the trainable parameters. Besides, the BN layer collects the moving average values as pair of mean and variance of $\gamma$ and $\beta$ during training to capture the global data statistics, and employ these estimated values for feature normalization in the testing phase. 

In addition, we have added a global averaging pooling layer after the extracted high-level features for compact semantic embeddings, which helps to significantly decrease the parameters of output dense layers (i.e., by 12 times specific in this network architecture) for alleviating overfitting issues.

\subsubsection{Learning strategy redesign}
The CT images used in this study present notable appearance differences for COVID-19 patients across different severity. For examples as shown in Fig.~\ref{fig:samples}, the mild patient may only contain a small lesion while severe patient can be infected almost in whole lung scope. Such large variance within the input space further presents difficulties for the model to explore a robust optimal solution from heterogeneous COVID-19 datasets. 
To address this problem, we expect a smooth learning process to facilitate the model optimization to reach a relatively robust solution. To this end, we propose to improve the COVID-Net learning strategy by adjusting learning rate more smoothly in a cosine annealing manner~\cite{loshchilov2016sgdr}.
Specifically, denoting the total training epoch as $T$, the learning rate at a current epoch $t$ is calculated as follows:
\begin{equation}
\centering 
\eta_t=\eta_{min}+\frac{1}{2}(\eta-\eta_{min})(1+\cos{(\frac{t}{T}\pi)}),
\end{equation}
\noindent where $\eta$ is the initial learning rate, $\eta_{min}$ is a predefined threshold of minimum learning rate. 

\subsection{Joint Learning Scheme with Redesigned COVID-Net}
Given insufficiency of COVID-19 samples from individual hospitals, it is usually desired to aggregate cases from different data sources for deep learning model development. On top of the redesigned COVID-Net backbone, we further propose a joint learning scheme to explicitly tackle the data heterogeneity problem for boosted diagnosis performance.
\subsubsection{Separate batch normalization at data heterogeneity}
Previous studies have revealed the limited improvement or even performance degradation of simple joint training at severe data heterogeneity~\cite{kevin2019baseline,liu2020msnet}. One crucial reason is that the BN layer in joint model will suffer from an inaccurate estimation of moving average values during the training phase due to the statistical difference across datasets (as shown in Fig.~\ref{fig:samples}). During testing phase, the estimated values cannot accurately represent the testing data statistics in each site and hence will lead to performance degradation. In this paper, we employ the domain-specific batch normalization (DSBN) method~\cite{liu2020msnet,chang2019domain,dou2020unpaired} by assigning an individual BN layer for each site independently to explicitly tackle the statistic discrepancy. 
As shown in Fig.~\ref{fig:framework}, we replace the BN layers incorporated at redesigned COVID-Net with the DSBN layers.
Compared with original BN layer, the DSBN layer enables to capture domain-specific moving values that can accurately represent the statistics of each site, also supplies domain-specific training variables of $\gamma$ and $\beta$ to tackle the inter-site variations by performing separate internal feature normalization.

\begin{table*}[t]
\renewcommand\arraystretch{1.15}
\centering
\caption{\small{Results of different methods on the two datasets for COVID-19 CT image classification (mean$\pm$std).}}
\vspace{2mm}
\scalebox{0.83}{
\begin{tabular}{p{2.8cm}|ccccc|ccccc}
\toprule[1pt]
\multirow{2}{*}{Methods} & \multicolumn{5}{c|}{Site A} & \multicolumn{5}{c}{Site B}\\ 

\cline{2-11}
&Accuracy & F1& Recall & Precision &AUC&Accuracy& F1 & Recall & Precision&AUC \\
\hline

Single (COVID-Net \cite{wang2020covid}) &77.12$\pm$0.98&76.03$\pm$1.13&70.97$\pm$2.37&80.04$\pm$2.87&84.08$\pm$0.92&63.12$\pm$2.09&61.09$\pm$1.28&57.73$\pm$2.94&64.03$\pm$3.91&71.09$\pm$2.18\\
Single (Redesign) &89.09$\pm$1.08&88.97$\pm$0.91&83.78$\pm$0.62&94.58$\pm$2.07&94.12$\pm$0.87&77.07$\pm$1.92&77.04$\pm$2.17&74.69$\pm$3.91&79.48$\pm$0.96&84.13$\pm$0.82\\

\hline
Joint (COVID-Net \cite{wang2020covid}) &68.72$\pm$1.94&69.17$\pm$1.93&69.41$\pm$3.91&68.27$\pm$1.21&74.78$\pm$2.91&63.27$\pm$2.82&59.78$\pm$3.12&54.19$\pm$4.17&64.27$\pm$3.81&68.12$\pm$2.11\\
Joint (Redesign)
&78.42$\pm$2.19&77.86$\pm$2.01&74.07$\pm$3.16&80.82$\pm$1.05&85.72$\pm$3.54&69.67$\pm$0.92&66.89$\pm$4.91&66.94$\pm$5.86&64.98$\pm$3.17&72.48$\pm$2.17\\
\hline

Series Adapter~\cite{rebuffi2017learning}
&85.73$\pm$0.71&86.19$\pm$1.65&81.91$\pm$2.61&90.98$\pm$0.79&92.93$\pm$1.42&70.01$\pm$3.82&67.08$\pm$3.09&74.91$\pm$1.89&63.04$\pm$4.87&73.92$\pm$2.36\\
Parallel Adapter~\cite{rebuffi2018efficient}
&82.13$\pm$1.91&82.39$\pm$1.78&80.02$\pm$2.47&83.51$\pm$1.87&89.99$\pm$0.97&74.93$\pm$1.83&73.46$\pm$1.68&71.81$\pm$2.47&79.84$\pm$1.75&80.29$\pm$1.76\\
MS-Net~\cite{liu2020msnet}
&87.98$\pm$1.31&88.73$\pm$1.20&84.91$\pm$2.83&93.78$\pm$2.76&94.37$\pm$0.79&76.23$\pm$1.81&76.54$\pm$1.73&74.07$\pm$1.29&79.29$\pm$1.48&82.19$\pm$1.47\\
\hline

SepNorm
&88.76$\pm$0.78&87.88$\pm$0.81&82.97$\pm$1.66&95.46$\pm$0.74&94.57$\pm$0.77&76.89$\pm$0.65&75.02$\pm$1.14&70.34$\pm$3.76&{\bf80.74$\pm$2.98}&83.94$\pm$0.43\\
+ Contrastive (\textbf{Ours})
&{\bf90.83$\pm$0.93}&{\bf90.87$\pm$1.29}&{\bf85.89$\pm$1.05}&{\bf95.75$\pm$0.43}&{\bf96.24$\pm$0.35}&{\bf78.69$\pm$1.54}&{\bf78.83$\pm$1.43}&{\bf79.71$\pm$1.42}&78.02$\pm$1.34&{\bf85.32$\pm$0.32}\\
\bottomrule[1pt]

\end{tabular}
}
\label{tab:jointsite}
\end{table*}

\subsubsection{Contrastive domain invariance enhancement}

In addition to tacking the inter-site heterogeneity under joint learning, we further aim to encourage robust semantic embeddings that cluster regardless of the data source domains. This is crucial, as the benefit from aggregating multi-site data would only be partially leveraged if the model fails to project inputs of different sites into a harmonized feature space. In this regard, we propose to explicitly promote the intra-class cohesion and inter-class separation of the semantic embeddings of infected (i.e., positive COVID-19) and non-infected cases across sites.

We adopt the contrastive learning~\cite{chen2020simple} to achieve that goal. 
Given a pair of samples $(m,n)$, we denote their semantic embeddings extracted after the global average pooling layer of the network as $e_m$ and $e_n$, which are  8096-dimensional vectors. In the preliminary experiment, we observed that imposing the compactness regularization directly on the semantic features might be a too strict constraint that impede the convergence. We therefore introduce an embedding network $H_\phi$ to project the embeddings to a lower-dimensional space.
The similarity between this pair of samples $(m,n)$ is then computed on the projected features instead of the original features as:

\begin{equation}
sim(m, n)=\frac{H_\phi{(e_m)} \cdot H_\phi{(e_n)}}{\parallel H_\phi{(e_m)} \parallel_2 \cdot \parallel H_\phi{(e_n)}\parallel_2}.
\end{equation}

We denote the pair $(m, n)$ as positive pair if sample $m$ and $n$ are of the same class, otherwise negative pair.
In each iteration, we randomly sample a minibatch of $K$ examples from the two sites. The contrastive loss over each positive pair ($m$, $n$) within the minibatch is defined as follows:

\begin{equation}
\small
\ell_{contrastive}(m, n)=-log\frac{exp(sim(m,n)/\tau)}{\sum_{k=1}^{K}\mathbb{F}(m, k) \cdot exp(sim(m, k)/\tau)},
\label{cont_loss}
\end{equation}
where the value of $\mathbb{F}(m,k)$ is 0 and 1 for positive and negative pair, respectively; $\tau$ denotes a temperature parameter. The final loss function is computed over all positive pairs in the given mini-batch for both $(m, n)$ and $(n, m)$. Trained in this way, the model will be enhanced to explore the domain invariance of representations such that the semantic embeddings of samples of same class can lie close to each other in angle space regardless of domain, and away from those of different classes.

\subsection{Overall Training Objective and Technical Details}
The overall training objective $\mathcal{L}_{overall}$ composes the cross entropy loss $\mathcal{L}_{ce}$ to assess the classification error and the contrastive loss $\mathcal{L}_{con}$ to regularize latent space:
 \begin{equation}
    \mathcal{L}_{overall}= \mathcal{L}_{ce} + \alpha \cdot \mathcal{L}_{con},
 \end{equation}
 
 \noindent where $\mathcal{L}_{ce} = \frac{1}{N}\sum_{i}-g_i \cdot \log{p_i}$, in which $N$ is the number of samples, $g_i$ denote the one-hot groundtruth label and $p_i$ is the predicted probability map, and the $\mathcal{L}_{con}$ sums over pairs according to Eq. \ref{cont_loss}.
The embedding network $H_\phi$ has two fully connected layers, with output size of 1024 and 128 using ReLU activation function. This component is only optimized with $\mathcal{L}_{con}$.

The framework is implemented with PyTorch \cite{paszke2019pytorch} using an Nvidia TITAN Xp GPU. The classification model and embedding network are trained from scratch with the same Adam Optimizer. The learning rate was initialized with 1e-4 and decayed with cosine annealing.
We have used grid search with a random small subset of the entire dataset to empirically adjust the hyper-parameters, setting the temperature parameter $\tau$ as 0.05 and $\alpha$ is 1.0.
For our proposed method and all the comparsion methods, we totally trained 100 epochs with batch size as 32, containing 16 images from each dataset. Considering the imbalance of sample number between the two datasets, we reloaded the smaller dataset by four times. Data augmentation of random crop and random vertical, horizontal flip were used to mitigate the overfitting problem.

\begin{comment}
\begin{table}[t]
\renewcommand\arraystretch{1.1}
\centering
\caption{Detailed sample numbers of data split of the two datasets.}
\vspace{2mm}
\scalebox{0.83}{
\begin{tabular}{c|cc|cc|c}
\toprule[1pt]
\multirow{2}{*}{Split}&\multicolumn{2}{c|}{SARS-CoV-2 Dataset}&\multicolumn{2}{c|}{COVID-CT Dataset }&\multirow{2}{*}{Total}\\
\cline{2-5}
&Non-infection &Infection &Non-infection &Infection\\
\hline
Train  & 923& 939 & 292 & 251  & 2948\\

Test  & 307& 313 & 105 & 98 & 823\\

Total & 1230 & 1252  & 397 & 349   & 3228\\
\bottomrule[1pt]
\end{tabular}}
\label{tab:data-split}
\end{table}
\end{comment}

\section{Experiments}
\label{sec:experiment}
\subsection{Datasets and Evaluation Metrics}
We adopt two public COVID-19 CT datasets to evaluate our joint learning framework, including \emph{SARS-CoV-2}~\cite{Soares2020sars} and \emph{COVID-CT}~\cite{zhao2020covid}.
To the best of our knowledge, these two datasets are the only relatively large-scale high-quality COVID-19 datasets which are currently publicly available for research.
Among the two datasets, the \emph{SARS-CoV-2} (denoted as Site A) consists of 2482 CT images from 120 patients, in which 1252 are positive with COVID-19 and 1230 are non-COVID but with other types of lung disease manifestations. The spatial sizes of these images range from $119 \times 104$ to $416 \times 512$.
The \emph{COVID-CT} dataset (denoted as Site B) includes 349 CT images from 216 patients containing clinical findings of COVID-19 and 397 CT images from 171 patients without COVID-19. Resolutions of these images range from $102 \times 137$ to $1853 \times 1485$.
For the preprocessing of the two datasets, all images are first resized to $224 \times 224$ in axial plane, and then normalized into zero mean and unit variance for intensity values along channel dimension.

Our experiment conducted four-fold cross-validation on the two datasets.
Following the literature of COVID-19 diagnosis~\cite{Soares2020sars}, we adopt five metrics to provide comprehensive evaluation for the models, including: (1) Accuracy (\%), (2) F1 score (\%), (3) Sensitivity (\%), (4) Precision (\%) and (5) AUC (\%). We report the results in form of average and standard deviation over three independent runs.

\begin{figure*}[!t]
\centering
\includegraphics[width=\linewidth]{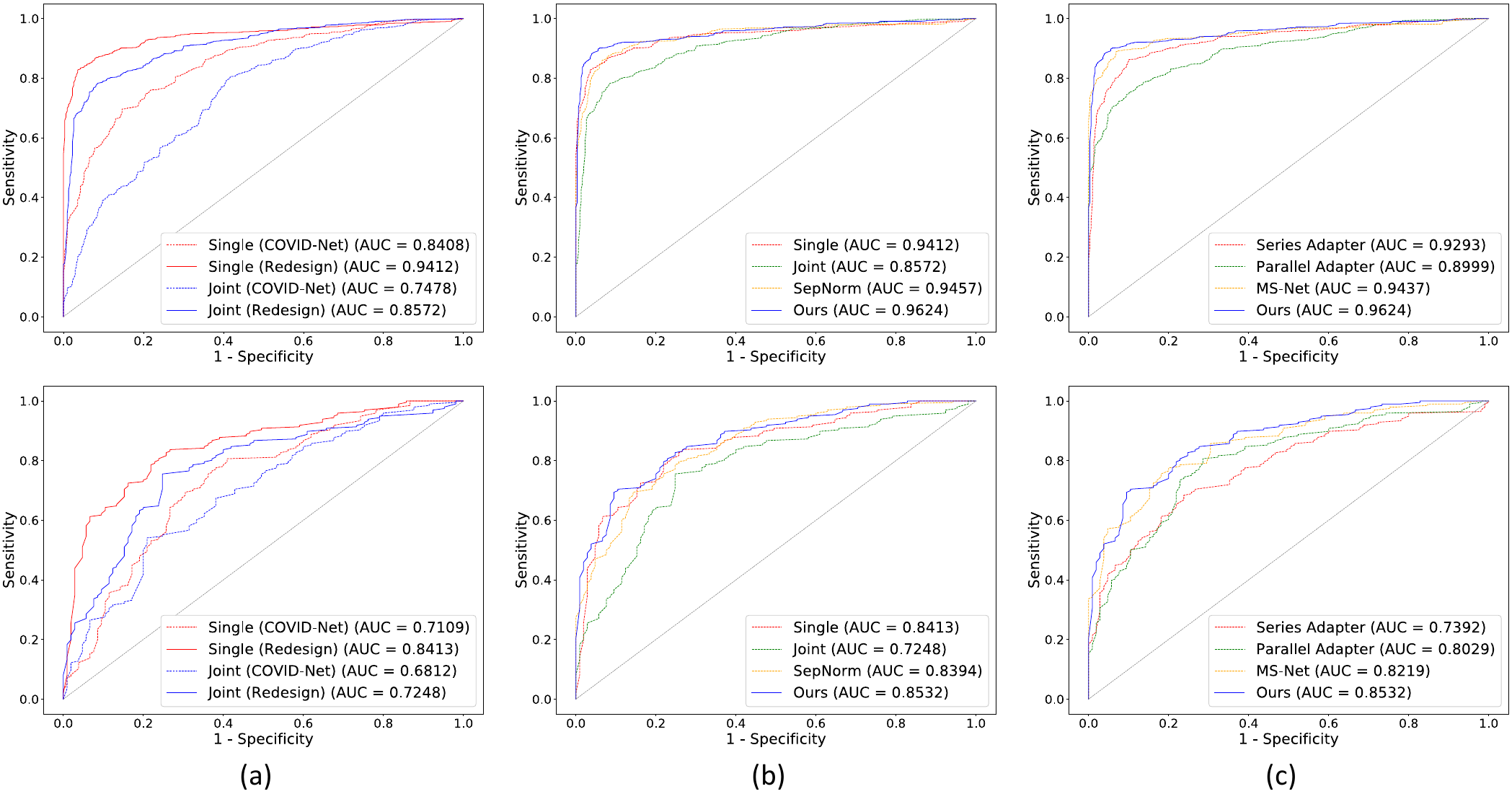}
\vspace{-3mm}
\caption{(a) ROC curves of Single and Joint approaches with redesigned and original backbone of COVID-Net on Site A (upper) and Site B (lower); (b) ROC curves of our approaches and baseline approaches (Single and Joint) on Site A (upper) and Site B (lower), using redesigned backbone; (c) ROC curves of our approach and other comparison methods on Site A (upper) and Site B (lower), using redesigned backbone.}
\label{fig:roc}
\end{figure*}

\subsection{Effectiveness of Network Redesign on COVID-Net}

We first compare our redesigned backbone with the original COVID-Net to validate the effectiveness of  network redesign. The comparisons are conducted on two different experimental settings, including 1) \emph{Single} setting which trains a model for each single site; and 2) \emph{Joint} setting which trains a model jointly using two datasets with naive aggregation.
From the results in Table~\ref{tab:jointsite}, we see that our \emph{Redesign} model outperforms the original COVID-Net \cite{wang2020covid} in \emph{Single} setting on both two sites by a large margin, with consistent increase on all five evaluation metrics. 
Similar observations are shown in \emph{Joint} setting, except the slightly marginal improvement of precision in Site B. 
These results highlight the superior representation learning ability of our redesigned backbone for COVID-19 diagnosis.
Fig.~\ref{fig:roc} (a) further displays the receiver operating characteristic (ROC) curves of the \emph{Single} and \emph{Joint} settings on the two sites, with our redesigned model and the original COVID-Net as backbone respectively. The benefits of our architecture and learning strategy redesign can be further observed from the overwhelming advantage in ROC curves.

\subsection{Effectiveness of Our Joint Learning Framework}

We then study the effectiveness of our proposed joint learning framework. Specifically, we first conduct comparison with the two baseline settings,~\ie, \emph{Single} and \emph{Joint}, and then compare with state-of-the-art joint learning approaches. Note that all these comparisons are based on the same backbone of redesigned COVID-Net for  fair comparison.  
\subsubsection{Comparison with baseline settings}
From the results in Table~\ref{tab:jointsite}, the \emph{Joint} approach underperforms the \emph{Single} approach in both Site A and Site B, with 8.40\% and 11.65\% decrease of AUC score respectively. Such performance degradation reveals the severe statistical discrepancy between the two datasets, and also highlights the urgency and clinical significance to design effective ways for improving the joint learning outcomes from heterogeneous datasets. It is worthy to point out that when conducting separate feature normalization for the two datasets, the joint learning model,~\ie, \emph{SepNorm}, outperforms the \emph{Joint} approach on both two sites consistently, which indicates the effectiveness of separate feature normalization scheme in solving the data heterogeneity problem. Notably, by further leveraging the proposed contrastive training objective, the model gains additional improvements on both Site A and Site B, achieving the AUC score of 96.24\% and 85.32\%, respectively. Such results demonstrate the effectiveness of the contrastive objective to promote more robust semantic embeddings from heterogeneous datasets. Our final results outperforms the \emph{Single} approach in 9 out of 10 metrics on the two sites, which further endorses the practical values of our approach to maximize the data utility of different datasets for boosting diagnosis accuracy. Fig.~\ref{fig:roc} (b) displays the ROC curves of our approach and the two baseline approaches for reference.

\begin{table}[!t]
\renewcommand\arraystretch{1.1}
\caption{\label{pvalue}The p-value with paired t-test of our method with Single, Joint and SepNorm learning schemes.}
\vspace{2mm}
\centering
\scalebox{0.9}{
\begin{tabular}{c|cccc}
\toprule[1pt]
Methods & Single & Joint & SepNorm \\
\hline
Site A & 0.002&0.007&0.023\\
Site B &0.004 &0.005&0.009\\
\bottomrule[1pt]
\end{tabular}
}
\end{table}

\begin{table}[!t]
\renewcommand\arraystretch{1.1}
\caption{The p-value with paired t-test of our method with the state-of-the-art comparison methods.}
\vspace{2mm}
\centering
\scalebox{0.9}{
\begin{tabular}{c|cccc}
\toprule[1pt]
Methods & Series-Adapter & Parallel-Adapter & MS-Net\\
\hline
Site A &0.009&0.012&0.021\\
Site B &1e-5&0.014&0.008\\
\bottomrule[1pt]
\end{tabular}
}
\label{pvalue2}
\end{table}
We conduct paired t-test to analyze the significance of the improvements of our method over the \emph{Joint}, \emph{Single}  and \emph{SepNorm} approaches.  The detailed results are shown in Table~\ref{pvalue}. We see that all paired t-tests present p-value smaller than 0.05, indicating the statistically significant improvements of our method on both two sites.

\subsubsection{Comparison with state-of-the-art methods}

We then compare our approach with state-of-the-art joint learning methods in both medical image analysis and natural imaging domain, including:
 \textbf{Series-Adapter}~\cite{rebuffi2017learning}:
    This study proposes series domain adapter for joint learning from multiple datasets, in which domain-adaptive layers are incorporated into residual block to mitigate the cross-domain visual discrepancy in natural image processing.
    \textbf{Parallel-Adapter}~\cite{rebuffi2018efficient}: They develop  parallel domain adapter where the domain-adaptive convolutional layer is inserted into residual block in parallel with filter banks to tackle the visual domain gap. This method achieves the state-of-the-art performance for the joint learning task from 10 different natural imaging classification datasets.
    \textbf{MS-Net}~\cite{liu2020msnet}: This work constructs a multi-site model that incorporates domain-specific auxiliary branches to improve the feature learning capacity and an online knowledge transfer strategy to explore the robust knowledge from multiple heterogeneous prostate MRI datasets for boosted segmentation.

The \emph{Joint} approach serves as a reference to evaluate these joint learning methods. As shown in Table~\ref{tab:jointsite}, the \emph{Series Adapter} achieves higher performance than \emph{Joint} model in both Site A and Site B, while its improvements are highly imbalanced across the two sites and still underperforms the \emph{Single} approach. Compared with \emph{Series Adapter}, the \emph{Parallel Adapter} presents relatively balanced improvements over the \emph{Joint} model, with 4.27\% and 7.81\% increase of AUC score in Site A and Site B respectively. 
Improvements of the two approaches over \emph{Joint} model indicate that the domain-specific parameters in domain adapter are beneficial for handling the problem of data heterogeneity. The \emph{MS-Net} is superior to the two domain-adaptive approaches, demonstrating the benefits of the knowledge transfer process in this framework. 
Notably, our method considerably outperforms all three state-of-the-art joint learning methods on both two sites, demonstrating the superiority of our approach to exploit more robust representations from heterogeneous datasets. The advantage of our method can also be reflected from the ROC curves in Fig.~\ref{fig:roc} (c). 
Results of paired t-test in Table~\ref{pvalue2} indicate the statistical significance of our improvements over the state-of-the-art methods.

\section{Discussions}
\label{sec:discussion}

\begin{figure}[t]
\centering
\includegraphics[width=\linewidth]{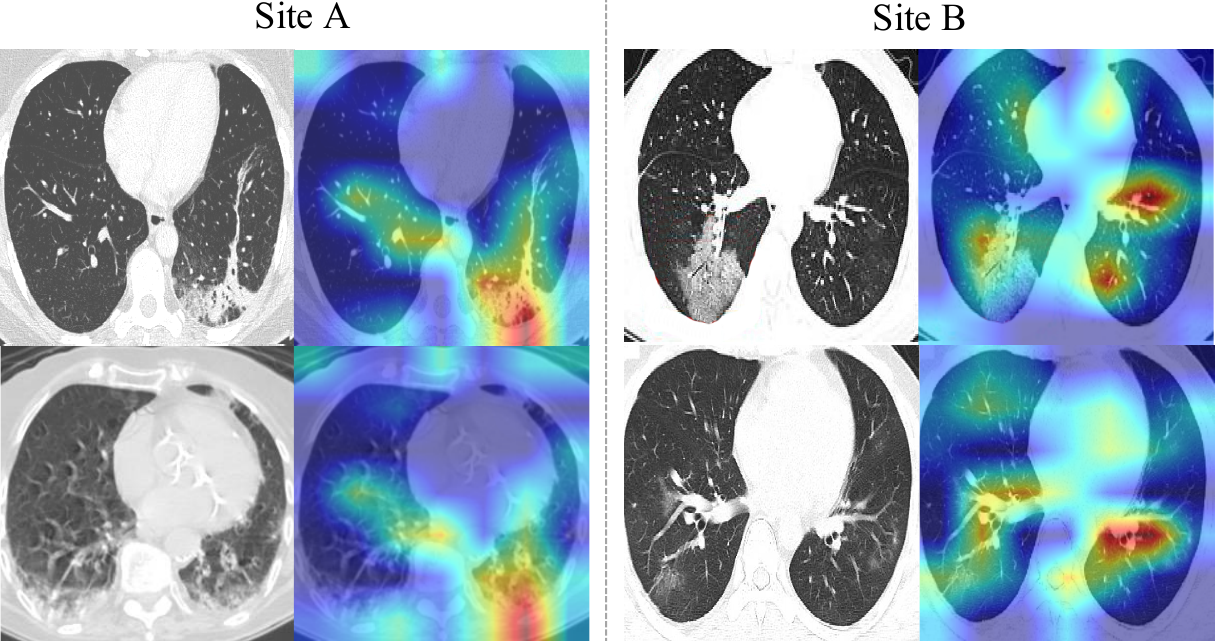}
\vspace{-5mm}
\caption{Visualization of color maps using Grad-CAM \cite{selvaraju2017grad}.}
\label{fig:gradcam}
\end{figure}

With  the  rapid  growth  rate  of  COVID-19  suspection all over the world, designing effective automated tools for COVID-19 diagnosis from CT imaging is highly demanded to improve the clinical diagnosis efficiency  and release the tedious workload of clinicians and radiologists. However, accurate diagnosis of COVID-19 from CT images is a non-trival problem, mainly due to the highly similar patterns of COVID-19 and other pneumonia types, as well as the large appearance variance of COVID-19 lesions of patients in different severity level~\cite{Ouyang2020}. 
Recently, a variety of data-driven models  have been proposed to solve this problem~\cite{fuhuazhu2020infnet, mobiny2020radiologist, gozes2020rapid, chaganti2020quantification}, leading to considerable progress in the field of automated COVID-19 diagnosis in the past few months. 

Appropriate network redesign is commonly required to adapt a well-established model onto a specific task. Our work employs the COVID-Net~\cite{wang2020covid} as backbone, which achieves superior performance in COVID-19 diagnosis with X-ray images than several popular classification networks. Considering that the CT images present more detailed and complex patterns of lesions than the X-rays, we redesign the COVID-Net in terms of network architecture and learning strategies to better capture the semantic representations and facilitate smooth learning process for boosted recognition  performance and learning efficiency on COVID-19 diagnosis from  CT images.

Given the large appearance variance of COVID-19 lesions and the highly similar patterns with other pneumonia types, the data-driven machine learning models certainly require a large-scale database for training to capture a widespread sample and lesion distribution to attain high accuracy~\cite{wang2020deep}. To mitigate the insufficiency of available COVID-19 CT scans from a certain hospital, it is meaningful and essential to collect the joint data efforts from different clinical centers for robust model development. Some previous studies have also highlighted the importance of learning from multi-site data for rapid and accurate model development in COVID-19 diagnosis~\cite{di2020hypergraph,javaheri2020covidctnet}, but most of them naively mix the data from different sources while ignoring the data heterogeneity that will affect the model to explore the general and robust knowledge for this task. Our experiment reveal that the separate feature normalization can effectively solve the problem of data discrepancy and the benefits of collaborative data efforts can be better explored by explicitly promoting the domain-invariant knowledge during training process.

\begin{figure}[t]
\centering
\includegraphics[width=\linewidth]{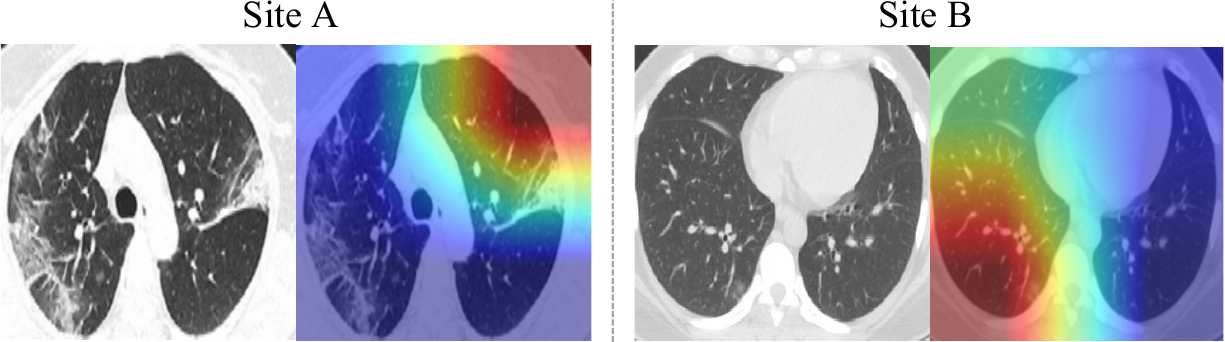}
\vspace{-5mm}
\caption{Visualization of color maps of failure cases with Grad-CAM~\cite{selvaraju2017grad}.}
\label{fig:failedcam}
\end{figure}

To understand the behavior of our framework, we observe the Grad-CAM \cite{selvaraju2017grad} visualization results on the two heterogeneous sites, as saliency maps (shown in Figure~\ref{fig:gradcam}). It is consistently observed on both datasets that the suspicious lesion regions are successfully localized across various abnormality patterns (e.g., bilateral and peripheral ground-glass, and consolidative pulmonary opacity), even with quite mild lesions. This analysis reveals promising interpretability of our classification model trained with image-level labels, demonstrating potential clinical relevance for COVID-19 image-based computer-assisted diagnosis.
In addition, we present typical failure cases in Figure~\ref{fig:failedcam}.
We see that the method would mis-classify samples due to wrongly attended regions, and fail to distinguish images with unobvious lesions. 

Although promising performance has been achieved as a preliminary study of multi-site learning with COVID-19 data, the limitation of our method still exists. Our method is limited to these two sites used in our paper, which is suboptimal to be directly applied on other unseen sites. This still cannot solve the challenge for wider cross-site deployment thoroughly.
Meanwhile, as the lack of computational resources and development urgency, we cannot pretrain our redesigned model on large-scale datasets such as ImageNet. Some previous works in the literature demonstrated that fine-tuning transferred models will bring performance improvement and speed up the training process~\cite{he2020sample}. 
As a near future work, we are interested to explore how to connect the carefully redesigned network architectures with model transfer learning from large-scale datasets, by trading off their respective benefits at balance.
In addition, we also plan to extend our method to more sites with different environments for wider multi-site learning to validate the generalization capability of AI models in the context of COVID-19 CT image diagnosis.

\section{Conclusion}
\label{sec:conclusion}
In this paper, we aim to develop a highly-accurate model for COVID-19 CT diagnosis by exploring the benefits of joint learning from heterogeneous datasets of different data sources. We propose a novel joint learning framework through redesigning the recently proposed COVID-Net from architecture and learning strategy as a strong backbone. Our joint learning framework explicitly mitigates the inter-site data heterogeneity by conducting separate feature normalization for each site. A contrastive training objective is further explored to enhance the learning of domain-invariant semantic features to improve the identification performance on each dataset. Experiments on two large-scale public datasets demonstrates the effectiveness and clinical significance of our approach. The future works include improving the generalization capacity of our model, extending it into a wider multi-site setting, as well as employing transfer learning from other large-scale datasets to further enhance the diagnosis accuracy.

\bibliographystyle{IEEEtran.bst}
\bibliography{refs.bib}

\begin{thebibliography}{10}
\providecommand{\url}[1]{#1}
\csname url@samestyle\endcsname
\providecommand{\newblock}{\relax}
\providecommand{\bibinfo}[2]{#2}
\providecommand{\BIBentrySTDinterwordspacing}{\spaceskip=0pt\relax}
\providecommand{\BIBentryALTinterwordstretchfactor}{4}
\providecommand{\BIBentryALTinterwordspacing}{\spaceskip=\fontdimen2\font plus
\BIBentryALTinterwordstretchfactor\fontdimen3\font minus
  \fontdimen4\font\relax}
\providecommand{\BIBforeignlanguage}[2]{{%
\expandafter\ifx\csname l@#1\endcsname\relax
\typeout{** WARNING: IEEEtran.bst: No hyphenation pattern has been}%
\typeout{** loaded for the language `#1'. Using the pattern for}%
\typeout{** the default language instead.}%
\else
\language=\csname l@#1\endcsname
\fi
#2}}
\providecommand{\BIBdecl}{\relax}
\BIBdecl

\bibitem{mei2020ai}
X.~Mei, H.-C. Lee, and e.~a. Diao, Kai-yue, ``Artificial intelligence–enabled
  rapid diagnosis of patients with covid-19,'' \emph{Nature Medicine}, 2020.

\bibitem{javaheri2020covidctnet}
T.~Javaheri, M.~Homayounfar, Z.~Amoozgar, R.~Reiazi, F.~Homayounieh, E.~Abbas,
  A.~Laali, A.~R. Radmard, M.~H. Gharib, S.~A.~J. Mousavi, O.~Ghaemi,
  R.~Babaei, H.~K. Mobin, M.~Hosseinzadeh, R.~Jahanban-Esfahlan, K.~Seidi,
  M.~K. Kalra, G.~Zhang, L.~T. Chitkushev, B.~Haibe-Kains, R.~Malekzadeh, and
  R.~Rawassizadeh, ``Covidctnet: An open-source deep learning approach to
  identify covid-19 using ct image,'' 2020.

\bibitem{xiayong2020covid}
J.~Zhang, Y.~Xie, Y.~Li, C.~Shen, and Y.~Xia, ``Covid-19 screening on chest
  x-ray images using deep learning based anomaly detection,'' in
  \emph{https://arxiv.org/abs/2003.12338}, 2020.

\bibitem{fuhuazhu2020infnet}
D.-P. Fan, T.~Zhou, G.-P. Ji, Y.~Zhou, G.~Chen, H.~Fu, J.~Shen, and L.~Shao,
  ``Inf-net: Automatic covid-19 lung infection segmentation from ct scans,'' in
  \emph{IEEE Transactions on Medical Imaging}, 2020.

\bibitem{di2020hypergraph}
D.~Di, F.~Shi, F.~Yan, L.~Xia, Z.~Mo, Z.~Ding, F.~Shan, S.~Li, Y.~Wei, Y.~Shao,
  M.~Han, Y.~Gao, H.~Sui, Y.~Gao, and D.~Shen, ``Hypergraph learning for
  identification of covid-19 with ct imaging,'' in
  \emph{https://arxiv.org/abs/2005.04043}, 2020.

\bibitem{wang2020covid}
L.~Wang and A.~Wong, ``Covid-net: A tailored deep convolutional neural network
  design for detection of covid-19 cases from chest x-ray images,'' 2020.

\bibitem{hu2018inter}
E.~Gibson, Y.~Hu, N.~Ghavami, H.~U. Ahmed, C.~Moore, M.~Emberton, and H.~J.
  Huisman, ``Inter-site variability in prostate segmentation accuracy using
  deep learning,'' in \emph{International Conference on Medical Image Computing
  and Computer-Assisted Intervention}, 2018.

\bibitem{john2018variable}
R.~Zech, John, A.~Badgeley, Marcus, M.~Liu, A.~B. Costa, J.~J. Titano, and
  E.~K. Oermann, ``Variable generalization performance of a deep learning model
  to detect pneumonia in chest radiographs: A cross-sectional study,''
  \emph{PLoS Med.}, vol.~15, no.~11, 2018.

\bibitem{liu2020msnet}
Q.~Liu, Q.~Dou, L.~Yu, and P.~A. Heng, ``Ms-net: Multi-site network for
  improving prostate segmentation with heterogeneous mri data,'' in \emph{IEEE
  Trans. Med. Imaging}, 2020.

\bibitem{shi2020review}
F.~Shi, J.~Wang, J.~Shi, Z.~Wu, Q.~Wang, Z.~Tang, K.~He, Y.~Shi, and D.~Shen,
  ``Review of artificial intelligence techniques in imaging data acquisition,
  segmentation and diagnosis for covid-19,'' \emph{IEEE Reviews in Biomedical
  Engineering}, 2020.

\bibitem{Butt_2020}
\BIBentryALTinterwordspacing
C.~Butt, J.~Gill, D.~Chun, and B.~A. Babu, ``Deep learning system to screen
  coronavirus disease 2019 pneumonia,'' \emph{Applied Intelligence}, Apr 2020.
  [Online]. Available: \url{http://dx.doi.org/10.1007/S10489-020-01714-3}
\BIBentrySTDinterwordspacing

\bibitem{hall2020finding}
L.~O. Hall, R.~Paul, D.~B. Goldgof, and G.~M. Goldgof, ``Finding covid-19 from
  chest x-rays using deep learning on a small dataset,'' \emph{arXiv preprint
  arXiv:2004.02060}, 2020.

\bibitem{narin2020automatic}
A.~Narin, C.~Kaya, and Z.~Pamuk, ``Automatic detection of coronavirus disease
  (covid-19) using x-ray images and deep convolutional neural networks,''
  \emph{arXiv preprint arXiv:2003.10849}, 2020.

\bibitem{abbas2020classification}
A.~Abbas, M.~M. Abdelsamea, and M.~M. Gaber, ``Classification of covid-19 in
  chest x-ray images using detrac deep convolutional neural network,''
  \emph{arXiv preprint arXiv:2003.13815}, 2020.

\bibitem{farooq2020covid}
M.~Farooq and A.~Hafeez, ``Covid-resnet: A deep learning framework for
  screening of covid19 from radiographs,'' \emph{arXiv preprint
  arXiv:2003.14395}, 2020.

\bibitem{li2020covid}
X.~Li and D.~Zhu, ``Covid-xpert: An ai powered population screening of covid-19
  cases using chest radiography images,'' \emph{arXiv preprint
  arXiv:2004.03042}, 2020.

\bibitem{Apostolopoulos}
\BIBentryALTinterwordspacing
I.~D. Apostolopoulos, S.~I. Aznaouridis, and M.~A. Tzani, ``Extracting possibly
  representative covid-19 biomarkers from x-ray images with deep learning
  approach and image data related to pulmonary diseases,'' \emph{Journal of
  Medical and Biological Engineering}, vol.~40, no.~3, p. 462–469, May 2020.
  [Online]. Available: \url{http://dx.doi.org/10.1007/s40846-020-00529-4}
\BIBentrySTDinterwordspacing

\bibitem{afshar2020covidcaps}
P.~Afshar, S.~Heidarian, F.~Naderkhani, A.~Oikonomou, K.~N. Plataniotis, and
  A.~Mohammadi, ``Covid-caps: A capsule network-based framework for
  identification of covid-19 cases from x-ray images,'' 2020.

\bibitem{gozes2020rapid}
O.~Gozes, M.~Frid-Adar, H.~Greenspan, P.~D. Browning, H.~Zhang, W.~Ji,
  A.~Bernheim, and E.~Siegel, ``Rapid ai development cycle for the coronavirus
  (covid-19) pandemic: Initial results for automated detection \& patient
  monitoring using deep learning ct image analysis,'' 2020.

\bibitem{tang2020severity}
Z.~Tang, W.~Zhao, X.~Xie, Z.~Zhong, F.~Shi, J.~Liu, and D.~Shen, ``Severity
  assessment of coronavirus disease 2019 (covid-19) using quantitative features
  from chest ct images,'' 2020.

\bibitem{Rahimzadeh2020}
\BIBentryALTinterwordspacing
M.~Rahimzadeh and A.~Attar, ``A modified deep convolutional neural network for
  detecting covid-19 and pneumonia from chest x-ray images based on the
  concatenation of xception and resnet50v2,'' \emph{Informatics in Medicine
  Unlocked}, vol.~19, p. 100360, 2020. [Online]. Available:
  \url{http://dx.doi.org/10.1016/j.imu.2020.100360}
\BIBentrySTDinterwordspacing

\bibitem{zhang2020clinically}
K.~Zhang, X.~Liu, J.~Shen, Z.~Li, Y.~Sang, X.~Wu, Y.~Zha, W.~Liang, C.~Wang,
  K.~Wang \emph{et~al.}, ``Clinically applicable ai system for accurate
  diagnosis, quantitative measurements, and prognosis of covid-19 pneumonia
  using computed tomography,'' \emph{Cell}, 2020.

\bibitem{DasAdhikari2020}
\BIBentryALTinterwordspacing
N.~C. Das~Adhikari, ``Infection severity detection of covid19 from x-rays and
  ct scans using artificial intelligence,'' \emph{International Journal of
  Computer (IJC)}, vol.~38, no.~1, pp. 73--92, May 2020. [Online]. Available:
  \url{https://ijcjournal.org/index.php/InternationalJournalOfComputer/article/view/1638}
\BIBentrySTDinterwordspacing

\bibitem{victor7effective}
U.~Victor, X.~Dong, X.~Li, P.~Obiomon, and L.~Qian, ``Effective covid-19
  screening using chest radiography images via deep learning,''
  \emph{Training}, vol.~7, no. 5,451, p. 152.

\bibitem{Nguyen_2019}
\BIBentryALTinterwordspacing
L.~H. Nguyen, J.~Zhu, Z.~Lin, H.~Du, Z.~Yang, W.~Guo, and F.~Jin,
  ``Spatial-temporal multi-task learning for within-field cotton yield
  prediction,'' \emph{Lecture Notes in Computer Science}, p. 343–354, 2019.
  [Online]. Available: \url{http://dx.doi.org/10.1007/978-3-030-16148-4_27}
\BIBentrySTDinterwordspacing

\bibitem{liao2018deep}
H.~Liao and J.~Luo, ``A deep multi-task learning approach to skin lesion
  classification,'' 2018.

\bibitem{glocker2019machine}
B.~Glocker, R.~Robinson, D.~C. Castro, Q.~Dou, and E.~Konukoglu, ``Machine
  learning with multi-site imaging data: An empirical study on the impact of
  scanner effects,'' 2019.

\bibitem{van2014transfer}
A.~Van~Opbroek, M.~A. Ikram, M.~W. Vernooij, and M.~De~Bruijne, ``Transfer
  learning improves supervised image segmentation across imaging protocols,''
  \emph{IEEE transactions on medical imaging}, vol.~34, no.~5, pp. 1018--1030,
  2014.

\bibitem{li2020multisite}
X.~Li, Y.~Gu, N.~Dvornek, L.~Staib, P.~Ventola, and J.~S. Duncan, ``Multi-site
  fmri analysis using privacy-preserving federated learning and domain
  adaptation: Abide results,'' 2020.

\bibitem{ioffe2015batch}
S.~Ioffe and C.~Szegedy, ``Batch normalization: Accelerating deep network
  training by reducing internal covariate shift,'' in
  \emph{https://arxiv.org/abs/1502.03167}, 2015.

\bibitem{loshchilov2016sgdr}
I.~Loshchilov and F.~Hutter, ``Sgdr: stochastic gradient descent with restarts.
  corr abs/1608.03983 (2016),'' \emph{arXiv preprint arXiv:1608.03983}, 2016.

\bibitem{kevin2019baseline}
L.~Yao, J.~Prosky, B.~Covington, and K.~Lyman, ``A strong baseline for domain
  adaptation and generalization in medical imaging,'' in \emph{Medical Imaging
  with Deep Learning-MIDL}, 2019.

\bibitem{chang2019domain}
W.-G. Chang, T.~You, S.~Seo, S.~Kwak, and B.~Han, ``Domain-specific batch
  normalization for unsupervised domain adaptation,'' in \emph{CVPR}, 2019.

\bibitem{dou2020unpaired}
Q.~Dou, Q.~Liu, P.~A. Heng, and B.~Glocker, ``Unpaired multi-modal segmentation
  via knowledge distillation,'' \emph{arXiv preprint arXiv:2001.03111}, 2020.

\bibitem{rebuffi2017learning}
S.-A. Rebuffi, H.~Bilen, and A.~Vedaldi, ``Learning multiple visual domains
  with residual adapters,'' in \emph{Advances in Neural Information Processing
  Systems}, 2017, pp. 506--516.

\bibitem{rebuffi2018efficient}
S.-A. Rebuffi, H.~Bilen, and A.~Vedaldi, ``Efficient parametrization of
  multi-domain deep neural networks,'' in \emph{Proceedings of the IEEE
  Conference on Computer Vision and Pattern Recognition}, 2018, pp. 8119--8127.

\bibitem{chen2020simple}
T.~Chen, S.~Kornblith, M.~Norouzi, and G.~Hinton, ``A simple framework for
  contrastive learning of visual representations,'' \emph{arXiv preprint
  arXiv:2002.05709}, 2020.

\bibitem{paszke2019pytorch}
A.~Paszke, S.~Gross, F.~Massa, A.~Lerer, J.~Bradbury, G.~Chanan, T.~Killeen,
  Z.~Lin, N.~Gimelshein, L.~Antiga \emph{et~al.}, ``Pytorch: An imperative
  style, high-performance deep learning library,'' in \emph{Advances in neural
  information processing systems}, 2019, pp. 8026--8037.

\bibitem{Soares2020sars}
E.~Soares, P.~Angelov, S.~Biaso, M.~Higa~Froes, and D.~Kanda~Abe, ``Sars-cov-2
  ct-scan dataset: A large dataset of real patients ct scans for sars-cov-2
  identification,'' \emph{medRxiv}, 2020.

\bibitem{zhao2020covid}
J.~Zhao, X.~He, X.~Yang, Y.~Zhang, S.~Zhang, and P.~Xie, ``Covid-ct-dataset: A
  ct scan dataset about covid-19,'' 2020.

\bibitem{selvaraju2017grad}
R.~R. Selvaraju, M.~Cogswell, A.~Das, R.~Vedantam, D.~Parikh, and D.~Batra,
  ``Grad-cam: Visual explanations from deep networks via gradient-based
  localization,'' in \emph{Proceedings of the IEEE international conference on
  computer vision}, 2017, pp. 618--626.

\bibitem{Ouyang2020}
\BIBentryALTinterwordspacing
X.~Ouyang, J.~Huo, L.~Xia, F.~Shan, J.~Liu, Z.~Mo, F.~Yan, Z.~Ding, Q.~Yang,
  B.~Song, and et~al., ``Dual-sampling attention network for diagnosis of
  covid-19 from community acquired pneumonia,'' \emph{IEEE Transactions on
  Medical Imaging}, p. 1–1, 2020. [Online]. Available:
  \url{http://dx.doi.org/10.1109/tmi.2020.2995508}
\BIBentrySTDinterwordspacing

\bibitem{mobiny2020radiologist}
A.~Mobiny, P.~A. Cicalese, S.~Zare, P.~Yuan, M.~Abavisani, C.~C. Wu, J.~Ahuja,
  P.~M. de~Groot, and H.~Van~Nguyen, ``Radiologist-level covid-19 detection
  using ct scans with detail-oriented capsule networks,'' \emph{arXiv preprint
  arXiv:2004.07407}, 2020.

\bibitem{chaganti2020quantification}
S.~Chaganti, A.~Balachandran, G.~Chabin, S.~Cohen, T.~Flohr, B.~Georgescu,
  P.~Grenier, S.~Grbic, S.~Liu, F.~Mellot, N.~Murray, S.~Nicolaou, W.~Parker,
  T.~Re, P.~Sanelli, A.~W. Sauter, Z.~Xu, Y.~Yoo, V.~Ziebandt, and
  D.~Comaniciu, ``Quantification of tomographic patterns associated with
  covid-19 from chest ct,'' 2020.

\bibitem{wang2020deep}
S.~Wang, B.~Kang, J.~Ma, X.~Zeng, M.~Xiao, J.~Guo, M.~Cai, J.~Yang, Y.~Li,
  X.~Meng, and B.~Xu, ``A deep learning algorithm using ct images to screen for
  corona virus disease (covid-19),''
  \emph{https://www.medrxiv.org/content/10.1101/2020.02.14.20023028v5}, 2020.

\bibitem{he2020sample}
X.~He, X.~Yang, S.~Zhang, J.~Zhao, Y.~Zhang, E.~Xing, and P.~Xie,
  ``Sample-efficient deep learning for covid-19 diagnosis based on ct scans,''
  \emph{medRxiv}, 2020.

\end{thebibliography}

\end{document}